\documentclass[draftcls,onecolumn,twoside,12pt]{IEEEtran}

\usepackage{graphicx}
\usepackage{bbm}

\usepackage{amsfonts,amssymb,amsmath}

\def\foorp{{\hfill$\spadesuit$}}

\def\inv{{^{-1}}}

\def\RR{{\mathbb R}}

\def\be{\begin{equation}}
\def\beq#1{\begin{equation}\label{#1}}
\def\ee{\end{equation}}
\def\bea{\begin{eqnarray}}
\def\beqa#1{\begin{eqnarray}\label{#1}}
\def\eea{\end{eqnarray}}
\def\ba{\begin{array}}
\def\ea{\end{array}}

\DeclareMathAlphabet{\mathpzc}{OT1}{pzc}{m}{it}

\def\cB{{\mathcal B}}

\def\cH{{\mathcal H}}

\def\cU{{\mathcal U}}
\def\cV{{\mathcal V}}

\def\Ltwo{L^2(\RR )}

\def\bds{_{-\infty}^\infty}
\def\supp{\hbox{supp}}
\def\mod{{\mathrm {mod}}}

\def\ba{{\mathbf a}}

\def\tu{{\tilde{u}}}
\def\tv{{\tilde{v}}}

\usepackage{type1cm,eso-pic,color}

\usepackage{mathpazo}
\usepackage[scaled=.95]{helvet}
\usepackage{courier}

\newtheorem{definition}{Definition}
\newtheorem{theorem}{Theorem}
\newtheorem{proposition}{Proposition}
\newtheorem{lemma}{Lemma}
\newtheorem{corollary}{Corollary}
\newtheorem{remark}{Remark}
\graphicspath{{figures/}}

\def\bmu{{\boldsymbol\mu}}
\def\bnu{{\boldsymbol\nu}}
\def\brho{{\boldsymbol\rho}}
\def\bsig{{\boldsymbol\sigma}}
\def\supp{{\rm supp}}

\def\tcU{{\widetilde{\cU}}}
\def\tcV{{\widetilde{\cV}}}

\def\beq{\begin{equation}}
\def\eeq{\end{equation}}

\def\defeq{\stackrel{\Delta}{=}}

\title{Refined support and entropic uncertainty inequalities}
\author{B. Ricaud\footnote{Supported by UNLocX, project FET-Open 255931}\hskip3mm
and B. Torr\'esani,~\IEEEmembership{Member,~IEEE,}\\
\normalsize
LATP, Aix-Marseille Univ. and CNRS, \\
\small
CMI, 39 rue Joliot-Curie, 13453
  Marseille cedex 13, France}
\date{\today}

\begin{document}
\maketitle

\begin{abstract}
Generalized versions of the entropic (Hirschman-Beckner) and support (Elad-Bruckstein) uncertainty principle are presented for frames representations. Moreover, a sharpened version of the support inequality has been obtained by introducing a generalization of the coherence. In the finite dimensional case and under certain conditions, minimizers of this inequalities are given as constant functions on their support. In addition, $\ell^p$-norms inequalities are introduced as byproducts of the entropic inequalities.
\end{abstract}


\section{Introduction}
The uncertainty principle is originally a quantum physics principle stating
that some families of observable quantities cannot be measured simultaneously
with infinite precision. The uncertainty principle can be turned into
quantitative statements thanks to uncertainty inequalities, which provide bounds
on precision of simultaneous measurements of such quantities.

The prototype of uncertainty inequality is the celebrated Heisenberg
inequality, first formulated in~\cite{Heisenberg27uber},
which uses a variance measure as criterion for the measurement
precision. Namely, for all $x\in\Ltwo$,
$$
\int\bds t^2 |x(t)|^2\,dt\,.\,\int\bds \nu^2 |\hat
x(\nu)|^2\,d\nu\ge\frac1{16\pi^2}\ ,
$$
where the Fourier transform is normalized in such a way that the Fourier
transform of $e^{-\pi t^2}$ equals $e^{-\pi\nu^2}$.
Originally stated for position and momentum, the Heisenberg
inequality has been extended to more general observable pairs, under the name of
Robertson inequality~\cite{Robertson29uncertainty,Robertson34indeterminacy}.
Particular cases have been analyzed by various authors, see
e.g.~\cite{Dahlke95affine,dahlke08uncertainty,Flandrin01inequalities,Maass10do}
and references therein. The Robertson variance inequality has been criticized in
the physics literature, mainly because the bound in the inequality sometimes
depends explicitely on the left hand side, which has motivated to seek
alternative formulations. Besides, Robertson-type inequalities do not generalize
well to all situations: for example, the notion of variance is not necessarily
easy to define in some contexts, such as for periodic sequences or functions,
functions defined on compact manifolds or graphs, more generally in situations
where the notion of spreading away from a reference point is not
straightforward. 
Among the generalizations, entropic inequalities, that use entropy measures to
quantize measurement precision have enjoyed renewed interest recently. In the
particular case of the 
position-momentum situation, the corresponding entropic uncertainty inequality,
called the Hirschman-Beckner inequality~\cite{Hirschman57note}, is intimately
related to the sharp form of the Hausdorff-Young inequality, the so-called
Babenko-Beckner inequality~\cite{Beckner75inequalities}.
In signal processing terms, this uncertainty principle limits the simultaneous
concentration or sparsity of a function and its Fourier transform. The
inequality provides a lower bound on the differential entropies of their
respective square moduli.

\medskip
Uncertainty inequalities have received a renewed interest in the context of
sparse approximation and signal processing applications. Often in a
finite-dimensional setting, $\ell^p$ norms (with $p<2$) are used to measure
dispersion of signals. This provides some quantities in order to compare the sharpness of different representations ($\ell^2$ vectors) of a signal $x$ or probe the concentration of information inside them. For example, the signal itself and its Fourier transform are two representations of the same mathematical object. More generally, any projection of $x$ on a basis of the Hilbert space gives a representation of the signal. In this context, uncertainty bounds involving $\ell^0$ quasi-norm and $\ell^1$ norm have
been derived. A prototype of such bounds is the Elad-Bruckstein $\ell^0$
inequality: given two orthonormal bases in a
finite-dimensional Hilbert space and any vector $x$ in that space with set of
coefficients $a$ and $b$ with respect to the two bases,
$$
\|a\|_0\,\|b\|_0\ge \frac{1}{\bmu^2}\ ,
$$
where $\bmu$ is a constant called mutual coherence, that depends on the two
bases (and not on $x$). Such results have important implications for practical problems, as shown
in the pioneering work of Donoho and Stark~\cite{Donoho89uncertainty}.
For instance, $\ell^0$ bounds have been used to prove the
equivalence of $\ell^0$ and $\ell^1$-based sparse recovery algorithms, under
suitable sparsity assumptions~\cite{Donoho01uncertainty,Elad02Generalized}.
Results of similar nature have also been obtained in the context of the Fourier
transform on abelian groups (see for
example~\cite{Tao05uncertainty,Krahmer08uncertainty,Murty12uncertainty}).
As is well known in information theory, and remarked also
in~\cite{Przebinda03three}, Shannon entropy and $\ell^p$ norms are closely
connected, through R\'enyi entropies. Inequalities involving R\'enyi
entropies~\cite{Maassen90discrete,Dembo91information}
actually imply both Shannon entropy inequalities and $\ell^p$ inequalities.

This study has been divided into two parts (Sec.~\ref{sec:lzero} and Sec.~\ref{sec:entropy}). In the first part, we analyze such support ($\ell^0$) inequalities in the context of frame
decompositions, as follows. Given two frames $\cU$ and $\cV$ in a Hilbert space
$\cH$, denote by $U$ and $V$ the corresponding analysis operator. $Ux$ and $Vx$ are the two representations of $x$ with respect to the frames. For any
$x\in\cH$, we prove for example a bound of the form
$$
\|Ux\|_0  \|Vx\|_0\ge \frac1{\bmu_*^2}\ ,
$$
where $\bmu_*$ is a constant that only depends on the two frames. 
In the case of orthonormal bases, these inequalities yield refined forms for
support inequalities ($\bmu_*\le\bmu$), for which we can analyze conditions for equality. The refined
inequalities involve cumulated coherence measures, instead of the standard
coherence measures used classically. In the case of frame decompositions, the
inequalities we obtain concern analysis coefficients, while most recent
contributions (in the domain of sparse decompositions and approximation) focus
on inequalities involving synthesis coefficients. Therefore, exact recovery
results such as those derived in~\cite{Elad02Generalized,Ghobber11uncertainty}
do not apply directly to the new results.
Though, given the renewed interest on analysis-based sparse
decompositions and co-sparsity (see e.g.~\cite{Nam11cosparse}), we believe that
these new inequalities are of interest, as they can yield bounds for the
performances of cosparse signal recovery methods. Consider for instance the
following signal separation problem: given two frames $\cU$ and $\cV$, and some
observed signal $u\in\cH$, we want to split $u$ as a sum of two components whose
respective analysis coefficients with respect to frames $\cU$ and $\cV$ are sparse.
In other words, we want to solve
$$
\min_{x,y\in\cH} \left[\|Ux\|_0 +\|Vy\|_0\right]
\quad\hbox{under constraint}\quad
u=x+y\ .
$$
where $U$ and $V$ are the analysis operators of two frames under consideration.
Given two such decompositions $u=x+y = x'+y'$, the above support inequality
directly leads to
\begin{align*}
\|Ux\|_0 + \|Vy\|_0 + &\|Ux'\|_0 + \|Vy'\|_0 \\
&\ge \|U(x-x')\|_0 + \|V(y'-y)\|_0\ge\frac2{\bmu_*}\ .
\end{align*}
Therefore, if one is given a splitting of the form $u=x+y$ such that
$\|Ux\|_0 + \|Vy\|_0 < 1/\bmu_*$, this splitting is automatically the solution
of the above optimization problem.

Besides support size estimates, we also obtain entropic inequalities
for analysis coefficients with respect to frames, that explicitely involve
the frame bounds. This is developed in the second part of the study. As a particular case, Shannon entropy bounds are derived, and
it is shown that the latter are only informative for tight frames.
In the latter case, the entropy inequalities take a fairly simple form; for
example, denoting by $S(a)$ the Shannon entropy of a vector $a$, we show that
given two tight frames $\cU$ and $\cV$, with respective analysis operators
$U$ and $V$, then
$$
S(Ux) + S(Vx)\ge -2\ln(\bmu_*)\ .
$$
Such inequality also turns out to yield the above mentioned support inequalities
as a by-product. Finally, we also derive new $\ell^p$ inequalities as consequences of
R\'enyi entropic inequalities.

\section{Refined Elad-Bruckstein $\ell^0$ inequalities}\label{sec:lzero}
\subsection{Notations}
We first introduce the general setting we shall be working with. Throughout this
paper, we shall denote by $\cU=\{u_k,\,k\in\Lambda\}$ and
$\cV =\{v_\ell,\,\ell\in\Lambda\}$ two countable frames for the Hilbert
space $\cH$ (we refer to~\cite{Christensen03Introduction} for a self contained
account of frame theory). Here, the index set $\Lambda$ will be finite when $\cH$ is
finite-dimensional, and infinite otherwise. $\|x\|$ also written $\|x\|_2$ is the norm of $x$ in $\cH$. We denote by $A_\cU,B_\cU$ and
$A_\cV,B_\cV$ the corresponding frame bounds, i.e. we have for all $x\in\cH$
\begin{align}
A_\cU\|x\|^2&\le\sum_k |\langle x,u_k\rangle|^2\le B_\cU\|x\|^2\ ,\\
A_\cV\|x\|^2&\le\sum_\ell |\langle x,v_\ell\rangle|^2\le B_\cV\|x\|^2\ .
\end{align}
Let $U:\cH\to\ell^2(\Lambda)$ and $V:\cH\to\ell^2(\Lambda)$ be the
corresponding analysis operators, i.e.
\beq
a_k\defeq(Ux)_k = \langle x,u_k\rangle\ ,\quad
b_\ell\defeq(Vx)_\ell = \langle x,v_\ell\rangle\ ,\qquad x\in\cH\ ,
\eeq
and denote by $T=VU^\dagger: a\to b$ the change of frame operator (with
$U^\dagger$ the Moore-Penrose pseudo-inverse of $U$).
We shall also denote by $\tcU$ and $\tcV$ corresponding (generic)
dual frames, among which the canonical dual frames will be denoted
by $\tcU^\circ = (U^*U)\inv\cU$ and $\tcV^\circ=(V^*V)\inv\cV$.
As is well known (see~\cite{Christensen03Introduction}), the corresponding frame
bounds are respectively $A_{\tcU^\circ}=1/B_\cU$, $B_{\tcU^\circ}=1/A_\cU$,
and similarly for $\tcV$. We recall that in the particular case where $\cU$
and/or $\cV$ are (Riesz) bases, $\tcU$ and/or $\tcV$ are the corresponding
bi-orthogonal bases.

In the following, we shall make use of the following quantity:
\begin{definition}
Let $r\in [1,2]$, let $r'$ be conjugate to $r$, i.e. such that $1/r + 1/r'=1$.
The mutual coherence of order $r$ of two frames $\cU$ and $\cV$ is defined by
\beq
\bmu_r(\cU,\cV) \defeq \sup_\ell \left(\sum_k |\langle
u_k,v_\ell\rangle|^{r'}\right)^{r/r'}\ ,
\eeq
In the case $r=1$, this corresponds to the standard mutual coherence, simply
denoted by $\bmu(\cU,\cV)$.
\end{definition}
This quantity is clearly well-defined in finite-dimensional settings. Notice
also that in infinite-dimensional situations (i.e. when $\Lambda$ is an
infinite index set), this quantity is well-defined for all $r\in [1,2]$. Indeed,
$\bmu_2(\cU,\cV)\le B_\cU\,\sup_\ell \|v_\ell\|^2$, and
$\bmu_r^{r'/r}(\cU,\cV)\le \bmu_2(\cU,\cV) \sup_{k,\ell}|\langle
u_k,v_\ell\rangle|^{r'-2}$, which is finite since $r'\ge 2$.

In finite-dimensional situations, the notion of mutually unbiased bases has been
introduced in the physics literature by Schwinger~\cite{Schwinger60unitary}
(see~\cite{Wehner10entropic} for a review).
\begin{definition}
Two orthonormal bases $\cU$ and $\cV$ in an $N$-dimensional Hilbert space $\cH$ are
mutually unbiased (MUB) if
$$
|\langle u_k,v_\ell\rangle| = \frac1{\sqrt{N}}\ ,\qquad\forall k,\ell=0,\dots
N-1\ .
$$
$\cU$ and $\cV$ are blockwise mutually unbiased bases (BMUB) of $\cH$ if
$\cU=\{\cU^{(1)},\dots\cU^{(K)}\}$, $\cV=\{\cV^{(1)},\dots\cV^{(K)}\}$, where
for all $k=1,\dots K$, $\cU^{(k)}$ and $\cV^{(k)}$ span the same subspace
$\cH^{(k)}$, of dimension $N_k$, and are MUBs for $\cH^{(k)}$, with
$\bigoplus_k\cH^{(k)}=\cH$.
\end{definition}
Notice that the coherence of a MUB in an $N$-dimensional Hilbert space equals
$\bmu(\cU,\cV)=1/\sqrt{N}$, the corresponding $r$-coherence equals
$\bmu_r(\cU,\cV)=N^{r/2-1}$, and the $r$-coherence of a BMUB equals
$\bmu_r(\cU,\cV)=\max_kN_k^{r/2-1}$.
\subsection{Refined Elad-Bruckstein inequality}
The classical Elad-Bruckstein $\ell^0$ inequality~\cite{Elad02Generalized} (a
strong form of which has been given in~\cite{Ghobber11uncertainty}) gives
a lower bound for the product of support sizes of two orthonormal basis
representations of a single vector. The inequality can be extended to the frame
case and generalized as follows.
\begin{theorem}
\label{th:ref.EB}
Let $\cU$ and $\cV$ be two frames of the Hilbert space $\cH$. For any $x\in\cH$,
$x\ne 0$, denote by $a=Ux$ and $b=Vx$ the analysis coefficients of $x$ with
respect to these two frames.
\begin{enumerate}
\item
For all $r\in[1,2]$, coefficients $a$ and $b$ satisfy the uncertainty inequality
\beq
\|a\|_0.\|b\|_0\ge\frac1{\bmu_r(\tcU,\cV)\bmu_r(\tcV,\cU)}\ .
\eeq
Therefore, $\|a\|_0.\|b\|_0\ge 1/{\bmu_*(\cU,\tcU,\cV,\tcV)}^{2}$, where
\beq
\label{fo:mustar}
\bmu_*(\cU,\tcU,\cV,\tcV)\defeq
\inf_{r\in [1,2]}\sqrt{\bmu_r(\tcU,\cV)\bmu_r(\tcV,\cU)}\ .
\eeq
\item
For all $r\in [1,2]$, the inequality can only be sharp if the following three
  properties hold true: 
\begin{itemize}
\item[i.] the sequences $|a|$ and $|b|$ are constant on their support,
\item[ii.] for all $k\in\supp(a)$ (resp. $\ell\in\supp(b)$) the sequence
$\ell\to|\langle\tu_k,v_\ell\rangle|$ (resp.
  $k\to|\langle\tv_\ell,u_k\rangle|$) is constant on $\supp(b)$ (resp. $\supp(a)$).
\item[iii.] for all $k\in\supp(a),\ell\in\supp(b)$,
$\arg(\langle\tu_k,v_\ell\rangle)=\arg(b_\ell)-\arg(a_k) =
  -\arg(\langle\tv_\ell,u_k\rangle)$. 
\end{itemize}
\end{enumerate}
\end{theorem}
\underline{\bf Proof:} 
\begin{enumerate}
\item
Let $r\in [1,2]$, let $x\in\cH$, $x\ne 0$. First remark that
\begin{eqnarray*}
\|b\|_\infty &=& \sup_\ell|\langle x,v_\ell\rangle|\\
&=&\sup_\ell \left|\left\langle\sum_k a_k \tu_k,v_\ell\right\rangle\right|\\
&\le& \sup_\ell \sum_k |a_k|\,|\langle\tu_k,v_\ell\rangle|\ ,
\end{eqnarray*}
and H\"older's inequality yields
\beq
\label{fo:frame.LrLinf.bound}
\|b\|_\infty\le \|a\|_r\ \bmu_r(\tcU,\cV)^{1/r}
\ee
Similarly,
\beq
\label{fo:frame.LrLinf.bound2}
\|a\|_\infty\le \|b\|_r\ \bmu_r(\tcV,\cU)^{1/r}\ .
\ee
Then, notice that
$$
\|a\|_r^r\le \|a\|_0\,\|a\|_\infty^r\le \|a\|_0\,\|b\|_r^r\ \bmu_r(\tcV,\cU)\ .
$$
The same estimate on $\|b\|_r$ proves the first part of the theorem.
\item
Assume first $r\ne 1$. As for the sharpness of the bound, notice first that the inequality
$\|a\|_r^r\le \|a\|_0\,\|a\|_\infty^r$ is sharp if and only if $|a|$ is constant
on its support (similarly, $|b|$ has to be constant on its support). Now in the
first inequality, H\"older's inequality is an equality if and only if the
sequence $k\to |\langle\tu_k,v_\ell\rangle|^{r'}$ is proportional to $|a|^r$,
meaning that the sequence  $k\to |\langle\tu_k,v_\ell\rangle|$ is constant on
its support, which coincides with the support of $a$. A similar reasoning is
done for $b$ and the sequence $\ell\to|\langle\tv_\ell,u_k\rangle|$. The last
inequality to be investigated is $|b_\ell| \le \sum_k
|a_k|\,|\langle\tu_k,v_\ell\rangle|$. The latter becomes an equality if and only
if the sum only involves positive numbers, i.e. iff
$\arg(\langle\tu_k,v_\ell\rangle)=\arg(b_\ell)-\arg(a_k)$. A similar reasoning
yields the condition $\arg(\langle\tv_\ell,u_k\rangle) =
\arg(a_k)-\arg(b_\ell)$.

Finally, consider the case $r=1$. The above argument can be reproduced exactly,
except for the tightness argument for H\"older's inequality. The latter can now
be an equality only if the sequence  $k\to |\langle\tu_k,v_\ell\rangle|$ is
equal to a constant (namely, $\bmu(\tcU,\cV)$) on $\supp(a)$, and smaller
outside the support. This does not change the conclusion.
\end{enumerate}
This concludes the proof.\foorp

\begin{remark}
\begin{enumerate}
\item
Clearly, by the arithmetic-geometric inequality, we also obtain the bound
\beq
\|a\|_0 + \|b\|_0\ge \frac2{\bmu_*(\cU,\tcU,\cV,\tcV)}
\eeq
\item
Using exactly the same techniques, the uncertainty inequality can be extended to
$K$ frames. Given $K$ frames $\cU^{(k)},\ k=1,\dots K$ and denoting by
$a^{(k)}$ the corresponding sequences of analysis coefficients of any $x\in\cH$,
we readily obtain the bound
\beq
\|a^{(1)}\|_0.\|a^{(2)}\|_0\dots\|a^{(K)}\|_0 \ge \left(\prod_{k=1}^{K}\mu_*^{(k)}\right)^{-1}
\eeq
where
$$\mu_*^{(k)}=\mu_*(\cU^{(k)},\tcU^{(k)},\cU^{(k+1 \mod K)},\tcU^{(k+1\mod K)}),
$$
and again by the arithmetic-geometric inequality,
\beq
\|a^{(1)}\|_0 +\|a^{(2)}\|_0 +\cdots+\|a^{(K)}\|_0\ge K\,
\left(\prod_{k=1}^{K}\mu_*^{(k)}\right)^{-1/K}\ .
\eeq
\end{enumerate}
\end{remark}
\begin{remark}
We notice that when $\cU$ and $\cV$ are orthonormal bases the result
generalizes the Elad-Bruckstein inequality. When $\cU$ and $\cV$ are
non-orthonormal bases $\tcU$ and $\tcV$ are their respective biorthogonal bases
and we obtain a straightforward generalization. In the case of frames, let us
point out that the generalization we obtain concerns analysis coefficients
rather than synthesis coefficients.
\end{remark}
\begin{remark}
Notice finally that these bounds involve arbitrary dual frames $\tcU$ and $\tcV$
of $\cU$ and $\cV$, not necessarily the canonical ones. Therefore the bound can
be make more general, in the form
\beq
\|a\|_0.\|b\|_0\ge 1/{\bmu_{**}(\cU,\cV)}^{2}\ ,
\eeq
where
\beq
\bmu_{**}(\cU,\cV)\defeq
\inf_{\tcU,\tcV}\ \inf_{r\in [1,2]}\sqrt{\bmu_r(\tcU,\cV)\bmu_r(\tcV,\cU)}\ ,
\eeq
the infimum running over the family of dual frames of $\cU$ and $\cV$. A
characterization of such families can be found
in~\cite{Christensen03Introduction}, Theorem 5.6.5.
\end{remark}
\subsection{Examples and comments: the case of orthonormal bases}
Consider first the case where $\cU$ and $\cV$ are two orthonormal bases in
finite dimensional Hilbert spaces. First
notice that the case $r=1$ provides an elementary proof of the Elad-Bruckstein
inequality (which involves $1/\bmu_1(\cU,\cV)^2$ as a lower bound), together
with explicit conditions for sharpness. In
the particular case of mutually unbiased bases, i.e. orthonormal bases such that
$|\langle u_k,v_\ell\rangle|$ is constant, $\bmu_r(\cU,\cV)=N^{r/2-1}$ is
monotone and minimal for $r=1$, which yields the usual coherence $\bmu =
\bmu_1=1/\sqrt{N}$, $N$ being the dimension of the considered Hilbert space. An
example of mutually unbiased bases is provided by the Kronecker and Fourier
bases in $\RR^N$, in which case the Elad-Bruckstein inequality coincides with
the inequality derived before by Donoho and Huo~\cite{Donoho01uncertainty}.
For blockwise mutually unbiased bases, we also obtain a monotone function of $r$
for the $r$-coherence $\bmu_r(\cU,\cV)=\max_kN_k^{r/2-1}$, which means that
again the optimal bound is provided by $\bmu_1$.

In the case of orthonormal bases, the smallest possible value for the coherence
is provided by the Welch bound: $\bmu\ge 1/\sqrt{N}$. Therefore, we obtain
\begin{corollary}
Assume $\cU$ and $\cV$ are orthonormal bases.
The optimal bound for the refined Elad-Bruckstein uncertainty inequality is
attained in the case of mutually unbiased bases, for $r=1$.
\end{corollary}
Consider now the case where the inequality is an equality, in the case $r\ne 1$.
By Theorem~\ref{th:ref.EB}, the analysis coefficients $a$ and $b$ of the
corresponding optimizer are such that $|a|$ and $|b|$ are constant on their
support. The proof of part 2. of the theorem also implies that for
$k\in\supp(a)$, the sequence $\ell\to|\langle u_k,v_\ell\rangle|$
vanishes outside $\supp(b)$ and equals a constant on $\supp(b)$; The latter
constant equals necessarily $\|b\|_0^{-1/2}$, and
$\bmu_r(\cU,\cV)=\|b\|_0^{r/2-1}$. Similarly, $\bmu_r(\cV,\cU)=\|a\|_0^{r/2-1}$.
Assume finally that the inequality be an equality, the latter thus reads
$$
\|a\|_0.\|b\|_0 = \|a\|_0^{1-r/2}.\|b\|_0^{1-r/2}\ ,
$$
which implies (for nonzero signals $x$) $\|a\|_0.\|b\|_0=1$, i.e. the two bases
have at least one common element, and the signal is a multiple of one of these
common elements.
\begin{corollary}
Assume $\cU$ and $\cV$ are orthonormal bases. For $r\ne 1$, the corresponding
refined Elad-Bruckstein inequality cannot be an equality, unless the two bases
have a common element.
\end{corollary}

\medskip

\begin{figure}
\centerline{
\includegraphics[width=7cm]{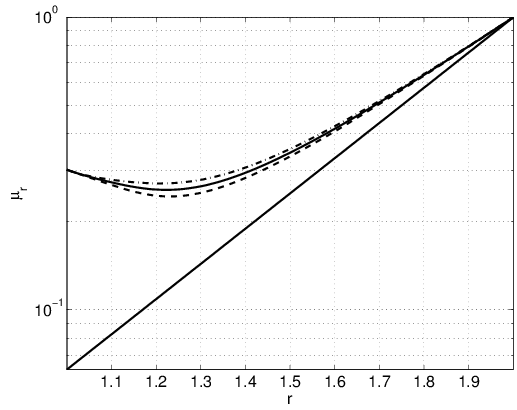}
}
\centerline{
\includegraphics[width=8cm]{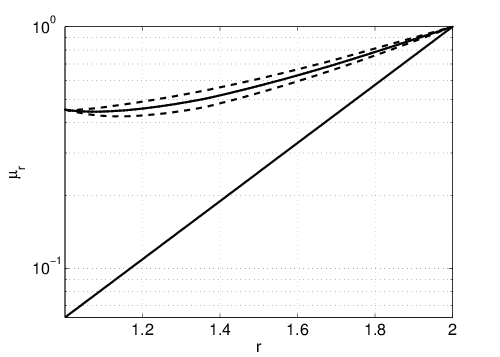}
}
\caption{Logarithm of $r$-coherence functions as a function of $r$.
$\bmu_r(\cU,\cV)$ (dashed), $\bmu_r(\cV,\cU)$ (dash-dotted) and
$\sqrt{\bmu_r(\cU,\cV)\bmu_r(\cV,\cU)}$ (full); straight  line: mutually
unbiased bases.
Top: random bases $\cU$ and $\cV$.
Bottom: MDCT bases with different window sizes.
}. 
\label{fi:rcoherence}
\end{figure}

Notice however that the case $r\in [1,2]$ constitutes a true generalization;
indeed, for general pairs of orthonormal bases, it turns out that
$\sup_r[1/\bmu_r(\cU,\cV)\bmu_r(\cV,\cU)] > 1/\bmu_1(\cU,\cV)^2$.
This is examplified in Figure~\ref{fi:rcoherence}, where are displayed the
functions $\bmu_r(\cU,\cV)$, $\bmu_r(\cV,\cU)$ and
$\sqrt{\bmu_r(\cU,\cV)\bmu_r(\cV,\cU)}$ as a function of $r$, in a generic
situation: the two bases $\cU$ and $\cV$ are random bases, obtained by
diagonalization of random (Gaussian) symmetric matrices. As can be seen in this
picture, the minimum of these three functions is not attained for $r=1$ but for
a larger value. For the sake of comparison, the case of mutually unbiased bases
is also represented and exhibit a power law behavior as a function of $r$
(represented as a straight line in the logarithmic plot). This shows that the
coherence based bounds are not the best possible ones in general. Elementary
infinitesimal calculus yields the following expression for the behavior of the
$r$-coherence near $r=2$:
\begin{align*}
\bmu_r(\cU,\cV) =& 1 \!-\! (2\!-\!r) \max_\ell \left(\!\!-\!\sum_k |\langle
u_k,v_\ell\rangle|^2 \ln\left(|\langle u_k,v_\ell\rangle|^2 \right)\!\right)\\
&+ O((2-r)^2)\ ,
\end{align*}
i.e. the slope of the tangent at $r=2$ is given by the entropy-like expression
$$
\hbox{slope}=-\sum_k|\langle u_k,v_\ell\rangle|^2\ln\left(|\langle u_k,v_\ell\rangle|^2
\right)
$$
(see section below), which is known to be minimal (in finite dimensional
situations, see~\cite{Cover91elements} for more details) when the $|\langle
u_k,v_\ell\rangle|^2$ are all equal.

More generally, we have the following result on $r$-coherences.
\begin{proposition}\label{prop:mur}
Let $\{u_k\}_k$ and $\{v_l\}_l$ be two frames in the Hilbert space $\cH$ of
dimension $N$.Fix $l$, let $s_l=\max_k|\langle u_k,v_l\rangle|$ and denote by
$n_l=\sharp (|\langle u_k,v_l\rangle|=s_l)$ the multiplicity of this maximal
value.
If $\max_l n_ls_l<1$, then there exists $r>1$ such that $\mu_r<\mu_1$.
\end{proposition}
\underline{\bf Proof:} 
It is enough to show that the derivative of $\mu_r$ is negative at $r=1$ under
the stated conditions. Let us introduce the notation  
$$
S_{kl}=|\langle u_k,v_l\rangle|\quad \text{ and }\quad L_l(r)=\ln \left(\sum_k
  S_{kl}^{r'}\right)^{r/r'}, 
$$
so that $\mu_r=\sup_l L_l(r)$. If for all $l$ the derivative of $L_l$ is
negative, so is the derivative of $\mu_r$.
Since $\{u_k\}_k$ and $\{v_l\}_l$ are frames, $\sum_k S_{kl}>0$ for all $l$ and
$L_l$ is well-defined as well as its derivative near $r=1$, $r\ge 1$. This
latter reads:
\begin{align*}
 L_l'(r)&=\ln n_ls_l+\sum_{k\in\Lambda}
 \alpha_{kl}^{\frac{r}{r-1}}+\frac{\ln s_l}{r-1}\sum_{k\in\Lambda}
 \alpha_{kl}^{\frac{r}{r-1}}\\
&+{\cal O}\left(\frac{1}{r-1}\left(\sum_{k\in\Lambda} \alpha_{kl}^{\frac{r}{r-1}}\right)^2\right),
\end{align*}
where $\alpha_{kl}=|\langle u_k,v_l\rangle|/n_ls_l$ and $\Lambda$ is the set of
$k$ such that $|\langle u_k,v_l\rangle|\neq s_l$.
For $r$ close to one, the dominant term is $\ln n_ls_l$. In this case, if
$\max_l n_ls_l<1$, the derivative of $\mu_r$ is negative.\foorp 
\begin{remark}
If two orthonormal bases have a high mutual coherence ($\mu_1$), this implies that a single term is dominant and since in this case $s_l\le 1$, Proposition~\ref{prop:mur} holds. However for frames with high coherence $s_l$ is large for some $l$, and it is highly probable that its multiplicity be large as well. In this case, the conditions of
the proposition are not satisfied and $\mu_r$ may not be smaller than
$\mu_1$. If there is too little coherence (like in MUB),
$n_l$ may be large and again $\mu_r$ may not be smaller than $\mu_1$. Finally, for MUB, $n_ls_l=\sqrt{N}$, this is the slope of the curve plotted on Fig.~\ref{fi:rcoherence}.
\end{remark}

\section{Entropic inequalities}\label{sec:entropy}
The support inequalities described above can also be obtained as particular
limits of entropic inequalities, which have been derived during the last 20
years in the mathematical physics and information theory communities. 
\subsection{Entropies}
In information theory, the notion of entropy is often used to measure disorder,
or information content of a random source; entropy measures are basically
related to measures of dispersion of the probability density function of the
random variables under consideration.

In the context of sparse analysis, the coefficients of the decomposition of any
finite-norm vector with respect to a frame can be turned into a probability
distribution, after suitable normalization. With the same notations as before,
denote by $a$ the sequence of analysis coefficients of $x\in\cH\ (x\ne 0)$ with respect to
the frame $\cU$, and we set $\tilde a = a/\|a\|_2$. Given
$\alpha\in[0,\infty]$ we introduce the corresponding R\'enyi entropy
\beq
R_\alpha(a) \defeq \frac1{1-\alpha}\,\ln\!\left(\|\tilde
a\|_{2\alpha}^{2\alpha}\right)\ .
\eeq
R\'enyi entropies fulfill a number of simple properties, among which we will use
the following two: monotonicity and limit to Shannon's entropy. More precisely,
for a given coefficient sequence $a$,
\beq
\alpha\le\beta\quad\implies\quad R_\alpha(a) \ge R_\beta(a)\ ,
\eeq
and
\beq
\lim_{\alpha\to 1} R_\alpha(a) = -\sum_n |\tilde
a_n|^2\ln\left(|\tilde a_n|^2\right)\defeq S(a)\ .
\eeq
$S(a)$ is the Shannon entropy of the coefficient sequence. Also, notice that
$R_0(a) = \ln \|a\|_0$. This will lead to support inequalities as consequences
of R\'enyi entropy inequalities.

Uncertainty inequalities involving entropy measures have been derived in several
different contexts
(see~\cite{Beckner75inequalities,Dembo91information,Maassen88generalized} for
example). We derive below similar inequalities in a more general setting.

\subsection{Entropic uncertainty inequalities for frame expansions}
As above, let us consider two frames $\cU$ and $\cV$. We use the same notations
as in the previous section, and introduce the following additional constants:
the geometric mean of the upper frame bounds $\brho$, the geometric mean of
frame bounds ratios $\bsig$, and the normalized $r$-coherence $\bnu_r$, written as
\begin{align}
\label{fo:notations}
&\brho(\cU,\cV)\defeq\sqrt{\frac{B_\cV}{A_\cU}}\ ,\qquad
\bsig(\cU,\cV) \defeq \sqrt{\frac{B_\cU B_\cV}{A_\cU A_\cV}}\ge 1\ ,\nonumber\\
&\bnu_r(\cU,\tcU,\cV) = \frac{\bmu_r(\tcU,\cV)}{\brho(\cU,\cV)^r}\ .
\end{align}
For the sake of simplicity, we shall drop the $r$ index in the case $r=1$, and
set $\bmu=\bmu_1$ and $\bnu=\bnu_1$.
We then have the following theorem, which can be seen as a frame generalization
of the Maassen-Uffink uncertainty
inequality~\cite{Maassen88generalized,Dembo91information}:
\begin{theorem}
\label{th:gen.entropic.ineq}
Let $\cH$ be a separable Hilbert space, let $\cU$ and $\cV$ be two frames of
$\cH$, and let $\tcU$ and $\tcV$ denote corresponding dual frames. Let
$r\in [1,2)$.
For all $\alpha\in [r/2,1]$, let $\beta=\alpha(r-2)/(r-2\alpha)\in [1,\infty]$.

For $x\in\cH$, denote by $a$ and $b$ the sequences of analysis coefficient of
$x$ with respect to $\cU$ and $\cV$. Then the R\'enyi entropies satisfy the
following bound:
\begin{eqnarray*}
(2-r) R_\alpha(a) + rR_\beta(b)
&\ge& -2\ln(\bnu_r(\cU,\tilde\cU,\cV))\\
&&- \frac{2r\beta}{\beta-1}\ln(\bsig(\cU,\cV))
\end{eqnarray*}
\end{theorem}
\underline{\bf Proof:} the proof is both a refinement and a frame
generalization of the proof in~\cite{Maassen88generalized,Dembo91information}. 
Let $T: a\to b$ denote the linear operator of change of coordinate.
From the frame bounds, we obviously have the inequalities
$$
\|b\|_2\le\sqrt{\frac{B_\cV}{A_\cU}} \|a\|_2\ ,\qquad
\|a\|_2\le\sqrt{\frac{B_\cU}{A_\cV}} \|b\|_2\ ,
$$
so that we have the estimate
\beq
\|T\|_{2\to 2} \le \sqrt{\frac{B_\cV}{A_\cU}} 
=\brho(\cU,\cV)\ .
\eeq
A second bound is obtained as in~\eqref{fo:frame.LrLinf.bound}, and yields
\beq
\|T\|_{r\to_\infty} = \bmu_r(\tilde\cU,\cV)^{1/r}\ .
\eeq
Let $p_0=q_0=2$, $p_1=r$,
$q_1=\infty$, and set for $\theta\in [0,1]$
$$
\frac1{p} = \frac{1-\theta}2 + \frac{\theta}r\ ,\qquad 
\frac1{q} = \frac{1-\theta}2\ .
$$
Clearly, $1-\theta = 2/q$ and $\theta = 1-2/q = r(1/p-1/q)$, and
the Riesz-Thorin lemma yields the following bound.
\beq
\|T\|_{p\to q}\le \bmu_r(\tcU,\cV)^{(1-2/q)/r}\brho(\cU,\cV)^{2/q}\ ,
\eeq
Using the definition of $\tilde a$ and $\tilde b$ and the frame bounds, we deduce
\begin{align}
\|\tilde b\|_q&\le
\brho(\cU,\cV)^{2/q}\bmu_r(\tcU,\cV)^{1/p -1/q}\sqrt{\frac{B_\cU}{A_\cV}}
\|\tilde a\|_p\nonumber\\
&\le\bsig(\cU,\cV)\bnu_r(\cU,\tcU,\cV)^{1/p-1/q}\|\tilde a\|_p\ ,
\end{align}
where we have used the bound $\|a\|_2/\|b\|_2\le\brho(\cV,\cU)$ and the
definition of $\bnu_r$ and $\bsig$ in~\eqref{fo:notations}.

Set now $p=2\alpha$ and $q=2\beta$; taking logarithms, we get
\begin{eqnarray*}
\frac{1-\alpha}{2\alpha} R_\alpha(a)-\frac{1-\beta}{2\beta} R_{\beta}(b)&\ge& 
-\left(\frac1{2\alpha}-\frac1{2\beta}\right)\,\ln(\bnu_r(\cU,\tcU,\cV))\\
&&\qquad- \ln(\bsig(\cU,\cV))\ ,
\end{eqnarray*}
Since  $(\beta-1)/\beta = 1-2/q = r(1/2\alpha-1/2\beta)$, this implies
\begin{eqnarray*}
\frac{\beta(1-\alpha)}{\alpha(\beta-1)}R_\alpha(a) + R_\beta(b) &\ge&
-\frac2{r}\ln\bnu_r(\cU,\tcU,\cV)\\
&&\qquad - 2\frac{\beta-1}\beta\ln(\bsig(\cU,\cV))\ .
\end{eqnarray*}

Finally, explicit calculations give $\alpha=\beta r/(r+2(\beta-1))$, so that
$$
\frac{\beta(1-\alpha)}{\alpha(\beta-1)} = \frac{2-r}r\in [0,1]\ ,
$$
which yields the desired result.\foorp

Notice that since $(2-r)/r\in [0,1]$, this implies the (generally non sharp)
inequality
$$
R_\alpha(a) + R_\beta(b)\ge
-\frac{2}r\ln(\bnu_r(\cU,\tcU,\cV)) -
\frac{2\beta}{\beta-1}\ln(\bsig(\cU,\cV))
$$
It is also worth noticing that in general, the limit $\alpha\to 1$ (which yields the
sum of the Shannon entropies as left hand side) is non-informative, since the
right hand side tends to $-\infty$, unless $\bsig=1$, i.e. $\cU$ and $\cV$ are
tight. In that case the following simplified inequalities hold true:
\begin{corollary}
Assume $\cU$ and $\cV$ are tight frames, and let $r\in[1,2)$:
\begin{enumerate}
\item
For all $\alpha\in [r/2,1]$, with $\beta=\alpha(r-2)/(2\alpha-r)\in [1,\infty]$
\beq
(2-r)R_\alpha(a) +r R_\beta(b)\ge -2\ln(\bnu_r(\cU,\tcU,\cV))\ .
\eeq
\item
the following inequalities between Shannon entropies hold true:
\beq
S(a)+S(b) \ge -2\ln\left(\bmu_*(\cU,\tcU,\cV,\tcV)\right)\ ,
\eeq
where $\bmu_*$ is defined in~\eqref{fo:mustar}.
\end{enumerate}
\end{corollary}
\underline{\bf Proof:} the first item is a direct consequence of the previous theorem in
the case of tight frames.
For the second item, remark that from the monotonicity of the R\'enyi entropy, we obtain
$(2-r)S(a) + rS(b)\ge (2-r)R_\alpha(a) +r R_\beta(b)$. Remark also that for tight
frames,
\begin{eqnarray*}
\bnu_r(\cU,\tcU,\cV)\bnu_r(\cV,\tcV,\cU) &=&
\frac{\bmu_r(\tcU,\cV)\bmu_r(\tcV,\cU)}{\bsig(\cU,\cV)^r}\\ &=&
\bmu_r(\tcU,\cV)\bmu_r(\tcV,\cU)\ .
\end{eqnarray*}
Symmetrizing the bound on Shannon entropies yield the desired result.
\foorp

\medskip
Notice that owing to the monotonicity property of R\'enyi entropies,
$R_0(a)=\ln(\|a\|_0)\ge S(a)$, and we recover the generalized Elad Bruckstein
inequality
$$
\|a\|_0.\|b\|_0\ge \frac1{\bmu_*(\cU,\tcU,\cV,\tcV)^2}\ .
$$

\medskip

Similar results in the general case are discussed below.

\subsection{Consequence: $\ell^p$ inequalities for analysis frame coefficients}
Let us start again from the modified entropic inequality in
Theorem~\ref{th:gen.entropic.ineq}, and symmetrize it with respect to $a$ and
$b$. We obtain
\begin{eqnarray*}
&(2-r)(R_\alpha(a)+R_\alpha(b)) + r(R_\beta(a)+R_\beta(b))\ge\\
&\qquad\qquad
-2\ln\left(\frac{\bmu_r(\tcU,\cV)\bmu_r(\tcV,\cU)}{\bsig(\cU,\cV)^r}\right)
-\frac{4r\beta}{\beta-1}\ln(\bsig(\cU,\cV))\ .
\end{eqnarray*}
Using the monotonicity of R\'enyi entropies, i.e. $R_\alpha\ge R_\beta$, we then
get for all $\alpha\in [r/2,1]$
\begin{eqnarray*}
R_{\alpha}(a) + R_{\alpha}(b) &\ge&
 -\ln\left(\bmu_r(\tcU,\cV)\bmu_r(\tcV,\cU)\right)\\
&&\qquad-r\frac{\beta+1}{\beta-1}\ln(\bsig(\cU,\cV))\ ,
\end{eqnarray*}
thus
\begin{eqnarray*}
\ln\left(\|\tilde a\|_{2\alpha}.\|\tilde b\|_{2\alpha}\right)\!\!\! &\ge&
-\frac{1-\alpha}{2\alpha}\ln\left(\bmu_r(\tcU,\cV)\bmu_r(\tcV,\cU)\right)\\
&&\ - r\frac{(\beta+1)(1-\alpha)}{2\alpha(\beta-1)}\ln(\bsig(\cU,\cV))\\
&\ge&\left(\frac1{2}-\frac1{2\alpha}\right)
\ln\left(\bmu_r(\tcU,\cV)\bmu_r(\tcV,\cU)\right)\\
&&\  -\left(1\!-\!\frac{r}2\right)\!
\left(1\!+\!\frac{r\!-\!2\alpha}{\alpha r\!-\!2\alpha}\right)\!\ln(\bsig(\cU,\cV))
\end{eqnarray*}
finally yields the bound, for $p\in [r,2]$
\begin{eqnarray}
\nonumber
\|a\|_p.\|b\|_p\! \!\!\! &\ge&
\!\!\!\!\left(\bmu_r(\tcU,\cV)\bmu_r(\tcV,\cU)\right)^{\frac1{2}-\frac1{p}}\\
&&\!\!\!\!\bsig(\cU,\cV)^{-(1-\frac{r}2)\left(1+\frac{r-p}p(1-\frac{r}2)\right)}
\|a\|_2.\|b\|_2\ .\quad
\end{eqnarray}
Also, using the fact that $R_0(a) =\ln(\|a\|_0)\ge R_\alpha(a)$ for all
$\alpha\in [r/2,1]$, and specifying to the sharpest bound $\alpha=r/2$,
we also obtain
$$
\ln\left(\|a\|_0.\|b\|_0\right)\ge
-\ln\left(\bmu_r(\tcU,\cV)\bmu_r(\tcV,\cU)\right)
-r\ln(\bsig(\cU,\cV))\ ,
$$
which yields
\be
\|a\|_0.\|b\|_0\ge\bsig(\cU,\cV)^{-r}\frac1{\bmu_r(\tcU,\cV)\bmu_r(\tcV,\cU)}\ .
\eeq
It is worth noticing that this bound is similar to the support inequalities
obtained previously, except for the factor $\bsig(\cU,\cV)^{-r}$, which makes it weaker.
Thus the bound is equivalent to the previous one if and only if the frames are
tight. Notice also that sharper bounds are obtained, as before, by optimizing
with respect to $r$ and the dual frames $\tcU$ and $\tcV$ of $\cU$ and $\cV$.

\subsection{Remark: necessary conditions for equality in the tight case}
We now examine conditions for the entropic inequalities be saturated. Our aim is
to make the connection with the {\em constant on support} property we already
met in Theorem~\ref{th:ref.EB} and its proof. Since the entropic bounds we could
prove are not sharp in generic situations, we limit the present discussion to the
particular case of tight frames. Let $\cU$ and $\cV$ be two tight frames,
denote by $A_\cU=B_\cU$ and $A_\cV=B_\cV$ the corresponding frame constants, and
set
\beq
g_{k\ell} = \langle u_\ell,v_k\rangle\ .
\eeq
Straightforward calculations give
$$
\frac{\partial}{\partial\overline{a}_\ell}\,\ln(\|a\|_{2\alpha}^{2\alpha}) =
\frac{\alpha}{\overline{a}_\ell}\,\frac{|a_\ell|^{2\alpha}}{\|a\|_{2\alpha}^{2\alpha}}
\ ,
$$
and
$$
\frac{\partial}{\partial\overline{a}_\ell}\,\ln(\|b\|_{2\beta}^{2\beta}) =
\sum_{k=0}^{N-1} \overline{g}_{k\ell} \frac{\beta}{\overline{b}_k}
\,\frac{|b_k|^{2\beta}}{\|b\|_{2\beta}^{2\beta}}
$$
and therefore the variational equations associated with the optimization of
$(2/r - 1)R_\alpha(a)+R_\beta(b)$ under constraint $\|x\|=1 = \|a\|/\sqrt{A}$ read
$$
\frac{2-r}r\frac{\alpha}{1-\alpha}\frac1{\overline{a}_\ell}
\frac{|a_\ell|^{2\alpha}}{\|a\|_{2\alpha}^{2\alpha}}
+ \frac{\beta}{1-\beta}\sum_{k=0}^{N-1} \overline{g}_{k\ell} \frac{1}{\overline{b}_k}
\,\frac{|b_k|^{2\beta}}{\|b\|_{2\beta}^{2\beta}} = \lambda
\frac{a_\ell}{A_\cU}\ ,
$$
where $\lambda$ is a Lagrange multiplier.
Now remark that $\beta/(1-\beta) = -\alpha(2-r)/r(1-\alpha)$; multiplying both
sides with $\overline{a}_k$ and summing over $k$, the constraint
$\|x\| = \|a\|/\sqrt{A_\cU}=1$ gives $\lambda=0$, so that the variational
equations take the form, for $\alpha\ne 1$
\beq
\frac{|a_\ell|^{2(\alpha-1)}}{\|a\|_{2\alpha}^{2\alpha}}\,a_\ell =
\frac1{A_\cU} \sum_{k=0}^{N-1} \overline{g}_{k\ell} 
\,\frac{|b_k|^{2(\beta-1)}}{\|b\|_{2\beta}^{2\beta}}\,b_k\ .
\eeq
\begin{remark}
From the above expression, we can observe that $|a|$ is constant on its support
if and only if $|b|$ is, since $\sum_k \overline{g}_{k\ell} b_k=a_l$. In this situation, we have $|a_k|=\sqrt{A_\cU/\|a\|_0}$
for all $k\in\supp(a)$ and  $|b_k|=\sqrt{A_\cV/\|b\|_0}$
for all $k\in\supp(b)$, so that
$$
R_\alpha(a) + R_\beta(b) = \ln\left(\|a\|_0.\|b\|_0\right)\ ,
$$
which therefore saturates the inequalities.
\end{remark}
Similar calculations on the Shannon entropy yield a comparable result.
\section{Conclusions}
We have examined in this paper entropic and $\ell^p$ uncertainty principles in
the framework of frame expansions. Our main results are extensions of support
and entropic uncertainty principles to the case of frames, which turn out to
generalize some known results when specializing to orthonormal bases. We showed
in particular that in general situations, bounds involving the classical mutual
coherence of the frames or bases under considerations are outperformed by the
new bounds involving generalized coherences.

While $\ell^p$ uncertainty principles have been mainly exploited in the
framework of sparse expansion problems, i.e. synthesis based approaches, our
results fit better into the so-called {\em analysis frameworks} (see
e.g.~\cite{Nam11cosparse}), as shortly explained in the introduction. Practical
consequences for co-sparse signal approximation and decomposition approaches are
still to be investigated further. This is ongoing work by the authors of the
present paper.

Let us mention that the finite dimensional case is by now fairly well
understood, and the existence of optimizers for the uncertainty inequalities is
closely connected to coefficient sequences that are constant on their support,
as already remarked by~\cite{Przebinda03three}.
In the infinite-dimensional case, such {\em constant on support} properties do
not make much sense in general situations, and the optimization problem is still
to be investigated much further.

\section*{Acknowledgements}
This work was supported by the UNLocX project of the Future Emerging
Technologies programme of the European Union (FET-Open grant number: 255931). B.
Torr\'esani also acknowledges partial support from the Metason project of the
french Agence Nationale de la Recherche CONTINT programme (ANR ANR-10-CORD-010).
\bibliographystyle{abbrv}
\bibliography{UnLocX,RicaudTorresani}

\end{document}